\documentclass[12pt]{iopart}
\usepackage{iopams}
\usepackage{graphicx}
\begin{document}
\title{Dynamic linear response of the SK spin glass coupled microscopically to a bath }
\author{T. Plefka}
\address{Theoretische Festk\"orperphysik, TU Darmstadt, D 64289 Darmstadt,
Germany} \ead{timm@arnold.fkp.physik.tu-darmstadt.de}
\begin{abstract}
The dynamic linear response theory of a general  Ising model
weakly  coupled to a heat bath is derived  employing the quantum
statistical theory of Mori, treating the Hamiltonian of the spin
bath coupling as a perturbation, and applying the Markovian
approximation. Both the dynamic susceptibility and the relaxation
function are expressed in terms of the static susceptibility and
the static internal field distribution function. For the special
case of the SK spin glass this internal field distribution can be
related to the solutions of the TAP equations in the entire
temperature region. Application of this new relation and the use
of  numerical solutions of the modified TAP equations  leads for
finite but large systems to explicit results for the distribution
function and for dynamic linear response functions. A detailed
discussion is presented which includes  finite-size effects. Due
to the derived temperature dependence of the Onsager-Casimir
coefficients a frequency-dependent shift of the cusp temperature
of the real part of the dynamic susceptibility is found.
\end{abstract}
\submitto{\JPA}
%\maketitle
\section{Introduction}
The Ising  model of Sherrington-Kirkpatrick (SK) \cite{sk} with
quenched random bonds is the most important  representative of a
class of long-ranged models all describing spin glasses (for
general references see  \cite{mpv,fh,young}). For the static
analysis of this model two complementary but conceptually
different approaches exist. The first approach uses the replica
method \cite{sk} and the breaking of the replica symmetry
\cite{par} to calculate bond-averaged quantities. The approach of
Thouless-Anderson-Palmer (TAP) \cite{tap}   is based on more
conventional techniques and does not perform this bond average as
it is expected that macroscopic physical quantities will be
independent of the particular configuration in the thermodynamic
limit.

Although the TAP equations are well established \cite{mpv,fh,I}
they are still a field of current interest. It is suspected that
not all aspects of this approach have yet been worked out.
Recently the author \cite{II} has reanalyzed the stability of
these equations and has shown that unstable states cannot be
described by the original TAP equations. Therefore he proposed
modified TAP equations which turned out to be useful for explicit
numerical calculations of the characteristic features of the SK
model of finite-sizes \cite{III}.

Dynamical questions are of great importance for the physics of
spin glasses (for reviews see \cite{mpv,fh,young}). Therefore
numerous  dynamical extensions have been added to the SK model.
Following the early dynamical approaches \cite{sk,kf} Glauber
dynamics has been used by various authors
\cite{II,glauber,szamel}. In addition, Langevin dynamics has been
employed in the studies \cite{langevin} mostly for the soft spin
version of the SK model. Both the Langevin and the Glauber
dynamics are basically phenomenological and can at best be
justified  partially by microscopic arguments. Note this also
applies to the work of Szamel \cite{szamel} although this approach
is  formulated in the `spirit' of the microscopic Mori theory.

It is the aim of this paper to present a  compete microscopic
analysis of the dynamical linear response for the SK model coupled
to a bath. An adequate tool for this purpose is the general theory
of Mori \cite{mori,fs}. This quantum-statistical  approach is used
in the present work and is worked out not only for the SK model
but for a general Ising model. Such a treatment is obvious and
straightforward. Nevertheless, to the authors knowledge it has
previously not been published .

As usual the results of the Mori theory are expressed in terms of
static equilibrium quantities. Together with the static isothermal
susceptibility, it is the internal field distribution function
\cite{ttcs,ttcss} which fully determines the dynamical linear
response. At this point we restrict the  approach to the SK model.
It is shown that the internal field distribution function can be
related to the solutions of the TAP equations. Employing the
approximate TAP solutions  \cite{III}, all the quantities of the
linear response theory can be numerically calculated in the entire
temperature regime for all external fields.

The references \cite{ttcs,ttcss} showed that an exact and complete
description of the thermodynamics of Ising models can be
formulated in terms of  the internal field distribution function.
Thus as a byproduct  the presented results of  this function are
of some interest independent of the dynamical linear response
problems.

Following a description of the microscopic Hamiltonian, the Mori
approach for a general Ising system is performed in section 2. The
internal magnetic field distribution function for the SK model is
treated in section 3. Both the  analytical and numerical results
for the field distribution function, the dynamic susceptibility
and the response function are explicitly presented in section 4.
Finally, some concluding remarks can be found in section 5.
\section{Linear response  for a general Ising system }
\subsection{ The microscopic system}
A system of  $N$ spins $ \bi{s}_i$ with $s={1\over2}$ is
considered in the presence of external fields $ H_i$.  The spins
interact via an  arbitrary Ising spin-spin interaction $
J_{ij}(=J_{ji})$ and are described by the spin Hamiltonian
\begin{equation}\label{1}
{\cal H}^S = - \frac{1}{2}
 \sum_{i,j}J_{ij}S_i S_j - \sum_i H_i S_i= - 2 \sum_{i,j}J_{ij}s_i^z s_j^z -2 \sum_i H_i s^z_i
\end{equation}
where $J_{ii}=0$ is presumed.  Both   quantum spin $1\over 2 $
operators $ s_i^z$ and  Ising spins $S_i(=2 s_i^z)$ are used
simultaneously in this work. Note that at this stage the bonds $
J_{ij}$ are quite general.

The assembly of spins is weakly coupled to a bath  described by
the Hamiltonian $ {\cal H}^B$ via a spin bath interaction
\begin{equation}\label{2}
{\cal H}^{S B}=  \sum_{i}\bi{B}_i \bi{s}_i= \sum_{i}\,\frac{1}{2}(
B^+_i s^-_i\,+\,B^-_i s^+_i)\,+\,B^z_i s^z_i
\end{equation}
where the operators $ \bi{B}_i$ represent variables of the bath
system. Thus the total Hamiltonian is given by
\begin{equation}\label{3}
{\cal H}={\cal H}^{0} +{\cal H}^{SB}\quad\textrm{with}\quad {\cal
H}^0={\cal H}^{B}+{\cal H}^{S}.
\end{equation}

There is no need for an explicit specification of the bath
Hamiltonian ${\cal H}^B $ and the bath operators $ \bi{B}_i$. As
shown below it is just the absorptive part $\chi_B''(\omega)$ of
the dynamic bath susceptibility
\begin{equation}\label{4}
 \chi_B(\omega)=\chi_B'(\omega) +i \chi_B''(\omega)=
-i\int_0^\infty \langle [\,B^-_i, e^{i \mathrm{L}^B\,t} B^+_i \,
]\rangle_{(B)}\; e^{ i\omega t} \mathrm{d}t \quad,
\end{equation}
which enters  the calculation and which is assumed to be known. In
equation (\ref{4})  $ \langle \ldots\rangle_{(B)}$ represents the
canonical thermal expectation value with respect to ${\cal H}^{
B}$ and $\mathrm{L}^B$ denotes the Liouville operator defined by
$\mathrm{L}^B A= [ {\cal H}^ B,A] $ . For later use it is noted
that $\chi_B''(\omega)$ may be rewritten as
\begin{equation}\label{5}
\chi_B''(\omega)= \pm \,\pi \,( e^{\pm \beta \omega}-1)\;\langle
B_i^\mp\,\delta ( \mathrm{L}^B\pm\omega)\,B_i^\pm \rangle _{(B)}
\end{equation}
which can easily be shown or can be found in literature \cite{fs}.
In writing equation (\ref{4}), it is assumed that the bath
susceptibilities do not depend on the site $i$. This assumption is
not essential and an extension to the general case is
straightforward.

Let $ \tau_B$ and $ \tau_S $  be the relaxation times of the bath
and the spin system respectively. Then it is natural to assume a
fast relaxation to thermal equilibrium for the bath system
compared to the spin system which  implies
\begin{equation}\label{6}
\tau_B\ll \tau_S \quad .
\end{equation}
For the  explicit determination of linear response quantities (see
section 4.)  the bath susceptibility $\chi_B''(\omega)$ enters for
$ \omega\tau_S\approx1 $. This implies $ \omega \tau_B\ll 1 $ and
\begin{equation}\label{7}
\chi_B''(\omega)\approx \textrm{const}\;
\omega\quad\textrm{for}\quad \omega \ll \tau_B^{-1}
\end{equation}
can be used in the generic case for the explicit calculations of
below. If  the simple form   $ \chi_B(\omega)=\chi_B ( 1-i \omega
/ \tau_B)$ is presumed, the constant of proportionality is given
by const$=\chi_B / \tau_B$, which may  in principle be temperature
dependent. This dependence  is neglected in this work by assuming
a slow variation on scale determined by the spin glass
temperature. This assumption is to a certain extent arbitrary but
can be justified for  special cases. Such a case is the Korringa
mechanism, where the bath and the $\bi{B}_i$ are identified as the
conduction electrons and itinerant spin density operators
respectively.
\subsection{Mori formalism  for the dynamic susceptibility }
The linear response of the  magnetizations
\begin{equation}\label{10}
\langle S_i\rangle (t)- \langle S_i\rangle= \sum_j
\chi_{ij}(\omega) h_j^{ex} \exp{i \omega t}\quad,\quad t\geq 0
\end{equation}
due to  small time dependent  external fields $
h_i^{ex}\,\Theta(t) \exp{ i\omega t}$ is governed by the dynamic
susceptibility matrix $\bchi(\omega)$ which is given by the Kubo
formula written in the Liouville space \cite{mori,fs}
\begin{equation}\label{11}
\chi_{ij}( \omega )= \beta\big( \tilde{ S}_i\big|\,\textrm{L}\,\{
\textrm{L}+\omega + i \eta \}^{-1}\big |\,\tilde{ S}_j\big) \quad
,\quad \eta \rightarrow + 0
\end{equation}
with $ \tilde{ A} =A -\langle A\rangle$. In the Liouville space
the operators $A$ of the state space are considered as vectors
$|A)$ with the temperature dependent Mori scalar product
\begin{equation}\label{13}
(A|B)= \frac{1}{\beta} \int_0^\beta \textrm{d}\lambda \langle
A^\dagger e^{-\lambda {\cal H}}Be^{\lambda {\cal H}}\rangle=
\frac{1}{\beta} \langle A^\dagger (1-e^{-\beta \mathrm{
L}})\textrm{L}^{-1} B\rangle
\end{equation}
where  $ \langle \ldots\rangle$ is the canonical thermal
expectation value with respect to ${\cal H}$. Operators in the
Liouville space like the Liouville operator
\begin{equation}\label{14}
\textrm{L}|A)=|[{\cal H},A])
\end{equation}
will be written in Roman letters \footnote{ The details of the
Mori approach including the justifications and the general
discussion of the approximations can be found in \cite{fs}. The
notation of the present work widely agrees with \cite{fs}.  Here
both the Boltzmann constant and $\hbar$ are set equal to $ 1$.}.

Introducing the projection operator in Liouville space
\begin{equation}\label{15}
\textrm{P}= \beta\sum_{ij}\big |\, \tilde{ S}_i \big)
\chi^{-1}_{ij}\big(\tilde{ S}_j\big |\quad \textrm{with} \quad
\chi_{ij} =\beta \big( \tilde{ S}_i\big |\,\tilde{  S}_j \big)
\end{equation}
and applying the standard Mori projection procedure \cite{mori,fs}
leads to
\begin{equation}\label{16}
\bchi^{-1}(\omega)=\bchi^{-1} -i\omega
\beta^{-1}\bi{L}^{-1}(\omega)\quad
\end{equation}
with
\begin{equation}\label{17}
\fl L_{ij}(\omega)=L'_{ij}(\omega)\,+i L''_{ij}(\omega)
 =i \big( S_i\big|\,\textrm{LQ}\,\{
\textrm{QLQ}+\omega + i \eta \}^{-1} \textrm{QL}\big|\, S_j\big)
\end{equation}
and with $\textrm{Q}=\textrm{1}-\textrm{P}$. The matrices $\bchi$
and $\bi{L}$ represent the static isothermal susceptibility matrix
and the dynamic Onsager-Casimir matrix respectively. Note that the
vectors $ \big |\,\tilde{  S}_i \big)$ which span the subspace $
\textrm{P}$ commute. Thus the frequency matrix vanishes in the
present case.

According to the standard approach \cite{fs} for a sufficiently
weak coupling ${\cal H}^{S B}$ the Markovian approximation
$\bi{L}(\omega)\approx \bi{L}'(0)$ can be applied in equation
(\ref{16}). Furthermore  the leading order perturbation-theory
expressions can be used for  both $\bchi$ and $\bi{L}'(0)$ in
(\ref{16}). The static isothermal susceptibility matrix is of the
order zero and approximated by
\begin{equation}\label{18}
\chi_{ij} =\beta \;\langle\tilde{ S}_i \,\tilde{
S}_j\rangle_{(S)}=\beta \;( \langle S_i \, S_j \rangle_{(S)}-m_i
m_j)
\end{equation}
where $ m_i=\langle S_i\rangle_{(S)} $ and where $ \langle
\ldots\rangle_{(S)} $ is the canonical thermal expectation value
with respect to the Ising Hamiltonian (\ref{1}).

Setting $\bi{L}=\bi{L}'(0)$ the lowest order of this matrix is the
second order in the perturbation and is given by
\begin{equation}\label{19}
\fl L_{ij}= \pi  \;\big( \textrm{Q}^{0}\textrm{L}^{S B}
S_i\;\big|\,\;\delta
(\;\textrm{Q}^{0}\textrm{L}^{0}\textrm{Q}^{0}\;)\;\textrm{Q}^{0}\textrm{L}^{SB}
S_j\big)_{(0)}=\pi  \;\big(\textrm{L}^{S B} S_i\;\big|\,\;\delta
(\,\textrm{L}^{0}\,)\;\textrm{L}^{SB} S_j\big)_{(0)}
\end{equation}
where $\textrm{L}^{SB}$ and $\textrm{L}^{0}$ are the Liouville
operators related to ${\cal H}^{S B}$ and to ${\cal H}^{0}$
respectively. The index $(0) $ denotes the Mori product taken with
${\cal H}^{0}$ alone and the projector $\textrm{Q}^{0}$ is defined
with the latter Mori product. The projector $\textrm{Q}^{0}$ drops
out, since $ S_i$ commutes with ${\cal H}^{0}$ and thus $
(\tilde{S}_i|\textrm{L}^{SB}S_j)=i \langle\, \tilde{S}_j\,[{\cal
H}^{S B},S_j\,]\rangle_{(0)}=0$ holds, where in addition  the
second equation of (\ref{13}) was used. Recalling  $J_{ii}=0$, one
finds that
\begin{equation}\label{20} \delta (\,\textrm{L}^{0}\,)\,B^\pm_i s^\mp_i= s^\mp_i\;\delta
\big(\,\textrm{L}^{B}\,\pm 2H_i\,\pm 2 X_i \big)\,B^\pm_i
\end{equation}
where the operator of the internal field $X_i$ at the site $i$ is
given by
\begin{equation}\label{201}
 X_i=\sum_j J_{ij} S_j \quad .
\end{equation}
Using equation (\ref{20}) we find  $ L_{ij}= L_{ii}\;\delta_{ij}$
with
\begin{equation}\label{21}
\fl L_{ii}=\frac{\pi}{2}\,\Big\langle (1-S_i)\; B^+_i\delta
(\textrm{L}^{B}-2H_i -2 X_i )\,B^-_i + (1+S_i)\;B^-_i\delta
(\textrm{L}^{B}+2H_i +2 X_i )\,B^+_i \Big\rangle_{(0)}\,.
\end{equation}
Let $O$ be any operator in the unitary space of the spins not
involving site $i$. Then
\begin{equation}\label{22}
\fl \langle \,S_i O\,\rangle_{(S)}=\Big\langle\;
\frac{\;\textrm{Tr}_i S_i O \exp( -\beta H_i-\beta
X_i)}{\textrm{Tr}_i \; \exp( -\beta H_i-\beta
X_i)}\Big\rangle_{(S)}= \langle \,O\, \tanh \beta (H_i +  X_i)
\,\rangle_{(S)}
\end{equation}
can be obtained \cite{ttcss}. Applying this relation to equation
(\ref{21}) and using (\ref{5}) finally yields for the dynamic
susceptibility matrix
\begin{equation}\label{231}
\fl \bchi^{-1}(\omega)=\bchi^{-1} -i\omega (\beta\bi{L})^{-1}\quad
\textrm{with}\quad L_{ij}=\delta_{ij}\, \,\Big \langle
\;\frac{\chi_B''( 2 H_i+2 X_i)}{\sinh \beta(2 H_i+2 X_i)}
\;\Big\rangle_{(S)}
\end{equation}
where the static isothermal susceptibility matrix $\bchi$ and the
internal field operators $ X_i$ are given by equation (\ref{18})
and by equation (\ref{201}) respectively.

To complete the analytic investigations for the general Ising
case,   the local-field probability distribution functions
\begin{equation}\label{26}
 P_i(h)\; = \langle \delta(h-H_i-X_i)\rangle_{(S)}=\langle \delta(h-H_i-\Sigma_j \;J_{ij}S_j)\rangle_{(S)}
\end{equation}
are introduced which permits the $L_{ii}$ to be rewritten as
\begin{equation}
\label{25} L_{ii}=\,\int \, P_i(h)\; \frac{\chi_B''(2 h)}{\sinh
(\beta 2 h)}\,\mathrm{d}h\quad .
\end{equation}
From the definition  the relations
\begin{equation}\label{27}\fl
\int P_i(h)\,\mathrm{d}h   = 1 \quad,\; \int h
\,P_i(h)\,\mathrm{d}h = H_i+\sum_j J_{ij}m_j \quad,\quad \int
\tanh(\beta h)\, P_i(h)\,\mathrm{d}h   =m_i
\end{equation}
immediately result, where the last relation is based on equation
(\ref{22}) with $O=1$.

With the result (\ref{231}) we have found in a very compact form
of the dynamic susceptibility matrix $ \bchi(\omega) $ . This
result is not restricted   to the SK model  and holds for all
Ising models. According to the equations (\ref{231}) and
(\ref{25}) the dynamic susceptibility $ \bchi(\omega) $ can be
explicitly calculated provided that the static susceptibility $
\bchi$ and the internal magnetic field distribution functions $
P_i(h)$ are known.

Knowledge of the linear dynamic susceptibility $ \bchi(\omega) $
implies  knowledge of all the  other response functions. Let us
consider the linear relaxation functions $ \bPhi(t)$  which
describe  the linear response $ \langle \tilde{S}_i\rangle (t)=
\sum_j [\Phi_{ij}(t)-\chi_{ij}]\; h_j^{ex} $ for $ t\geq 0$ due to
 small changes of the external fields $ -h^{ex}_j \Theta(t)$.
According to the general response theory $\bPhi (t)$ is given by
\begin{equation}\label{28}\Phi_{ij}(t)= \beta( \tilde{
S}_i\big|\,\tilde{ S}_j (t)\big)
\end{equation}
and is approximated by
\begin{equation}\label{29}
\frac{\textrm{d}\bPhi(t)}{\textrm{d}t}= - \beta\,
{\bi{L}}\bchi^{-1}\, {\bPhi}(t) \quad   \textrm{or by }\quad
\bPhi(t)=\bchi \,\exp( - \bchi^{-1} \beta \, {\bi{L}}\,t)\quad.
\end{equation}
The remaining quantity of interest in linear response theory is
the response function matrix which equals $-\dot{\bPhi}(t) $. Due
to this simple relation the response function matrix is not
further considered in this work.

The results (\ref{29}) can be  compared in detail with the work of
Szamel \cite{szamel} which represents both the most recent and the
closest treatment on the subject of this paper. Comparing equation
(\ref{29}) for the case $H_i=0$ with equation (13) of the first
paper of reference \cite{szamel} clearly shows that  the values $
(1 -\langle S_i \tanh \beta X_i \rangle_{(S)}) $ or according to
relation (\ref{22}) the values $ (1 -\langle \tanh^2 \beta X_i
\rangle_{(S)} )$ are used for the $ L_{ii}$ instead the correct
values  given in equation (\ref{231}). Thus in general both
approaches  disagree.

The work of Szamel  and other former work \cite{glauber} is based
on the phenomenological Glauber dynamics. Microscopic derivation
\cite{V,just} of this master-equation (in the form of reference
\cite{szamel}) leads  to the transition rates
\begin{equation}\label{2000}
w_i=( 1- S_i\tanh \beta X_i)\frac{\chi_B''( 2 X_i)}{4 \tanh \beta
X_i}\quad
\end{equation}
for the $H_i=0$ case. In the work of Szamel the values for the
rates $( 1- S_i\tanh \beta X_i)/2$ were used which can only be
justified for the special, rather unrealistic case that the bath
susceptibility $ \chi''_B(\omega)$ is proportional to $
\tanh(\beta \omega /2) $. Thus, for a general agreement with  the
present approach,  the correct transition rates (\ref{2000}) have
to be used in the master-equation treatment.

Microscopically  unjustified rates of the form of \cite{szamel}
are widely used to analyze the dynamics of Ising models for
various physical questions. Provided  the characteristic  width of
the distribution $P(h)$ is of the order of the temperature $T$ or
larger than $T$  a modified master-equation approach with the
rates (\ref{2000}) will lead to significant changes. Note that for
(standard) mean field treatments these effects are absent. In
these cases  $P(h)$ is a $\delta$-function and the modifications
reduce to a factor which can be eliminated by scaling  the time.
For the spin glasses at low temperatures, however, the exact form
of the transition rates is essential, as worked out in the
following for the SK model.
\section{Internal field distribution function for the SK spin glass}
For the rest of the paper we consider exclusively  the special
case of the SK model. In this case the bonds $ J_{ij}$ are
independent random variables with zero means and standard
deviations $N^{- {1\over 2}}$. This scaling fixes the spin glass
temperature to $T=1$. The smallness and the randomness enter
basically into the approach of this section where a tractable form
for the  field distribution  $ P_i(x) $ of the SK model is
deduced.

This approach uses  techniques similar to the derivations of the
TAP equations \cite{tap,mpv,II}. The spin Hamiltonian (\ref{1}) is
rewritten  as ${\cal H}^S=-(H_i+X_i)S_i +\hat{\cal H}_i$ where
$\hat{\cal H}_i$ describes a $ N-1$ spin system with the spin
$S_i$ removed from the original spin system. Using  the relation
$\delta(h-H_i-X_i) \exp[\beta(H_i+X_i)S_i]=\delta(h-H_i-X_i)
\exp(\beta h S_i)$ and the Fourier representation of the
$\delta$-function and taking the trace over the site $i$
\begin{equation} \label{40}\fl
 P_i(h)\;  =\frac{\cosh (\beta h)}{i\pi Z}
 \int_{-\infty}^{\infty}\hat{Z}_i(k)\,e^{-ik(h-H_i)} \,\textrm{d}k
 \quad\textrm{with}\quad \hat{Z}_i(k)= \hat{\Tr}_i \;e^{i k \sum_j
 J_{ij} S_j -\beta \hat{\cal H}_i}
\end{equation}
is obtained.  The partition function of the original spin system
is denoted by $ Z $  and $\hat{\Tr}_i$  stands for the trace of
the $N-1$ spin system.

The quantity $\hat{Z}_i(k)$ represents a partition function of the
$N-1$ spin system in presence of additional imaginary fields $ i k
\beta^{-1} J_{ij}$. Due to the smallness of the $ J_{ij}$ the
cumulant expansion can be performed which yields to leading order
in $N^{-1} $ \cite{tap,mpv,II}
\begin{equation}
\label{41}
 \ln \hat{Z}_i(k)= \ln \hat{Z}_i(0) +i k \sum_j J_{ij} \hat{m}_j
-\frac{k^2}{2 \beta} \sum _{j\,l} J_{ij}\hat{\chi}_{j l} J_{li}
\end{equation}
where $\hat{m}_j= \beta^{-1} \partial \ln \hat{Z}_i(0)/ \partial
H_j$ and $\hat{\chi}_{j l}=  \partial \hat{m}_j/
\partial H_{l}$ are the local magnetizations and the static
susceptibilities of the $N-1$ spin system respectively. Following
again the former approaches \cite{tap,mpv,II} the double sum in
equation (\ref{41}) is replaced by the local static susceptibility
\begin{equation} \label{42}
\chi_l=\frac{1}{N}\sum_j \chi_{jj}= \frac{1}{N}\sum_j
\frac{\partial m_j}{\partial H_j}\quad.
\end{equation}
With the result (\ref{41}) the integration in equation (\ref{40})
can be performed  yielding
\begin{equation} \label{43}\fl
P_i(h)= \sqrt{\frac{\beta}{2 \pi \chi_l}}\;\frac{\cosh (\beta
h)}{\cosh (\beta H_i^{eff})}\;\exp\Big\{-\frac{\beta
\chi_l}{2}\Big\}\; \exp\Big\{-\frac{\beta(h-H_i^{eff})^2}{2
\chi_l}\Big\}
\end{equation}
where the factors independent of $h$ are determined from the
normalization condition $1=\int P_i(h)\,\textrm{d}h $ and where
the local effective field $H _i^{eff} = H_i+\sum_jJ_{ij}\hat{m}_j$
was introduced. The latter two relations of equation (\ref{27})
yield
\begin{equation} \label{44}
H _i^{eff} = H_i+ \sum_jJ_{ij}m_j - m_j \chi_l \quad \textrm{and}
\quad m_i=\tanh (\beta H _i^{eff})
\end{equation}
which is employed  to rewrite equation (\ref{43}) as function of $
h, m_i$ and $\chi_l$
\begin{equation} \label{45}\fl
P_i(h)= \sqrt{\frac{\beta (1-m_i^2)}{2 \pi \chi_l}}\;\cosh (\beta
h)\;\exp\Big\{-\frac{\beta \chi_l}{2}\Big\}\;
\exp\Big\{-\frac{\beta(h-\beta^{-1}\mathrm{artanh}\: m_i)^2}{2
\chi_l}\Big\}\;.
\end{equation}

Note that equation (\ref{44}) are the well known TAP equations if
the local static susceptibility $ \chi_l$ is specified. In this
work the modified version \cite{II} of these equations is used
which yields
\begin{equation}\label{46}
\chi_l =\frac{1}{N}\sum_i
\frac{\beta(1-m_i^2)}{1+\Gamma^2\,\beta^{2}(1-m_i^2)^2}
\end{equation}
with
\begin{eqnarray}\label{47}\fl
\Gamma = 0\quad &\mathrm{f}&\mathrm{or}\quad
1-\frac{\beta^2}{N}\,\sum_i(1-m_i^2)^2\geq 0\\\fl
1=\frac{1}{N}\sum_i
\frac{\beta^2(1-m_i^2)^2}{1+\Gamma^2\,\beta^{2}(1-m_i^2)^2}\qquad
 &\mathrm{f}&\mathrm{or}\quad
1-\frac{\beta^2}{N}\,\sum_i(1-m_i^2)^2\leq 0\label{48}\quad.
\end{eqnarray}
As worked out in  \cite{II}, for the stable states  the modified
TAP equations are equivalent to the original ones in the
thermodynamic limit and differences result only  for unstable
states. For finite systems the complete temperature and field
dependence of the solutions of the modified equations is known
\cite{III}, in contrast to the original TAP equations.

With  the knowledge of these solutions the total internal field
distribution function defined as
\begin{equation}\label{49}
P(h) =\frac{1}{N}\sum_i\;P_i(h)
\end{equation}
can be calculated according to equation (\ref{43}) or (\ref{45}).
In zero field above the spin glass temperature  the  distribution
function reduces to
\begin{equation} \label{50}\fl
P(h)=P_i(h)= \frac{1}{\sqrt{2 \pi}}\;\cosh (\beta
h)\;\exp\Big\{-\frac{\beta^2+h^2}{2}\Big\}\quad \textrm{for}\quad
T\geq 1 \;,\quad H_i=0 \quad.
\end{equation}
The latter result for the special case has already  been obtained
by Thomsen \etal \cite{ttcss}. However, to the best of the authors
knowledge, the general results of this section have previously not
been published.
\section{Discussion and numerical results}
\subsection{Internal field distribution function}
The static properties of Ising models can entirely be formulated
in terms of the internal field distribution function $ P(h)$
\cite{ttcs,ttcss}. Due to this general importance  some discussion
of $ P(h)$ for the SK model is presented.

\begin{figure}
\begin{center}
\includegraphics{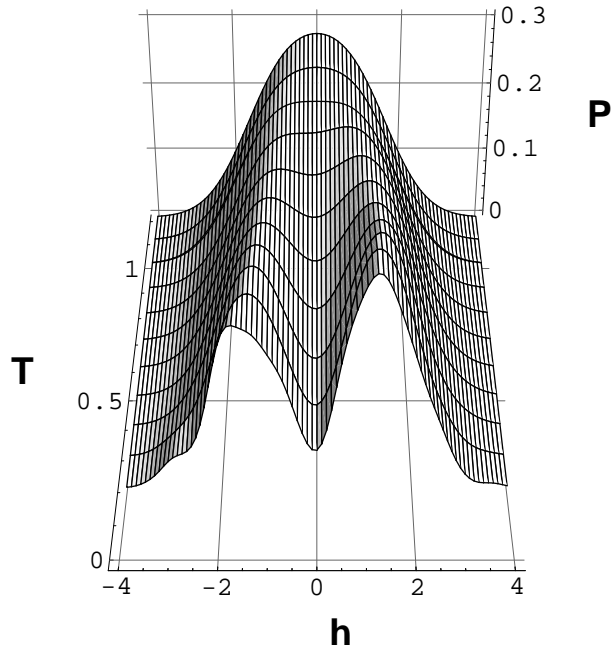}
\end{center}
\caption{\label{f1} Temperature dependence of the internal field
distribution function $ P(h)$ of the SK  spin glass for
temperatures $T $ below and above the spin-glass temperature $T=1$
for zero external field.}
\end{figure}
For zero magnetic field the temperature dependence of $ P(h) $ is
shown in figure (\ref{f1}). Above the spin- glass temperature
$T=1$ the exact result (\ref{50}) is plotted. Below the spin glass
temperature the figure is based on equation (\ref{45}) and on the
numerical results of \cite{III}. In the plot the  local
magnetizations $m_i$  for the state of lowest free energy of
 a   N=225 system (sample I) are used. On the scale of figure
(\ref{f1}) both regimes $ T\geq 1 $ and $ T\leq 1 $ fit smoothly
together. The distribution function $ P(h) $ bifurcates from a
one-peak structure to a two-peak structure at the spin-glass
temperature. With decreasing temperature the minimum located at $
h=0$ becomes deeper and finally reaches the value $P(h=0)=0$  for
zero temperature within numeric precision. This  behavior also
applies to the metastable states. Moreover the variations for the
different states are small and seem to become negligible in the
thermodynamic limit. This is remarkable  and is an indication for
self averaging of $P(h)$. These findings, as well as the general
temperature dependence of $P(h)$, are in agreement with
Monte-Carlo simulations of references \cite{ttcss}.
\begin{figure}
\begin{center}
\includegraphics{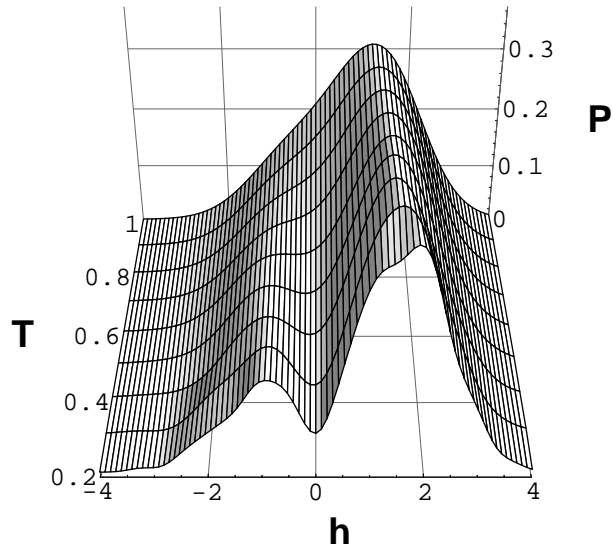}
\end{center}
\caption{\label{f2} Temperature dependence of internal field
distribution function $ P(h)$ of the SK  spin glass  in presence
of a homogenous external field H=0.5. The (numerical) AT
temperature is T=.577 .}
\end{figure}

The discussion is completed by figure (\ref{f2}) where $P(h)$ is
shown for a system  with a homogenous external field $ H=0.5$. No
exact results are known for this case and thus the plot is based
on numerical data everywhere. Again the data of sample I of
reference \cite{III} are used. Now the distribution $  P(h)$ is
asymmetric but again a bifurcation to a two-peak structure is
found when the spin glass regime is entered. This occurs at the
Almeida-Thouless (AT) temperature \cite{at} which is determined
from $ T^2=N^{-1} \sum_i (1-m_i^2)^2$ and which leads to the
numerical value of $ T= .577 $ for the sample under consideration.
These results  indicate that the spin glass regime is also
characterized by a two-peak structure of $ P(h)$  for the case
where a finite external field is present.

Certainly in the numerical results finite-size effects are
present. The most obvious feature is the asymmetry of $ P(h) $ in
figure (\ref{f1}). Thus the investigations of the finite-size
effects \cite{III} are extended to the distribution function
$P(h)$  and averages  over a few tens of independent samples are
performed keeping  $N $ and $T$ fixed. For the averages the
asymmetry of $P(h)$ for the case $H=0$ reduces both with
increasing  number of samples and with increasing $N$ . This
represents some numerical evidence that the asymmetry is indeed an
artifact due to the finite system size.

In this context we recall that in zero magnetic field for each
solution of the TAP equation (\ref{44}) a further solution
 can be constructed  trivially by changing the sign of all $m_i$.
According to equation (\ref{43}) the distributions corresponding
to these two solutions are given by $ P(h) $ and by $ P(-h) $
respectively. The means of these two distributions  exhibit
considerable smaller sample-to-sample variations than the
individual  $ P(h) $. This can in principle  be used to construct
improved  approximations of $ P(h) $ for the thermodynamic limit.
The TAP approach, however, does not use any averaging and thus we
avoid this procedure. Moreover the asymmetry of figure (\ref{f1})
indicates the order of magnitude of the  finite-size effects.
\subsection{Dynamic susceptibility}
We focus on the local dynamic  susceptibility which is defined as
\begin{equation}\label{80}\fl
\chi_l(\omega)= \chi_l'(\omega)+i\,\chi_l''(\omega)= N^{-1} \Tr
\bchi(\omega)=N^{-1} \Tr\big\{\bchi^{-1} - i \omega ( \beta {\bi
L})^{-1} \big\}^{-1}
\end{equation}
and which is  a quantity of  both theoretical and experimental
interest. Employing the  results of the last subsection and the
approximation (\ref{7}) (setting const=$1$ which fixes the time
scale) the Onsager coefficient $L_{ii}$ can  be numerically
determined according to equation (\ref{25}). Furthermore using the
well known expression for the static isothermal susceptibility
\begin{equation}\label{81}
\chi_{ij}^{-1}= \Big\{ [\beta(1-m_i^2)]^{-1}+
\chi_l\Big\}\;\delta_{ij} - J_{ij}
\end{equation}
with $ \chi_l$  given by the equations (\ref{46}-\ref{48}), all
terms of equation (\ref{231})  are explicitly known and the
dynamic susceptibility  matrix $\bchi(\omega) $ is obtained by
numerical matrix inversion. From this the local susceptibility
$\chi_l(\omega)$ is finally calculated.

A further method exists  to determine $\chi_l(\omega)$ which is
based on the theorem of Pastur \cite{pastur} \footnote{Actually a
slight generalization of this theorem is needed to incorporate the
term $i \omega \beta^{-1}\bi{L}^{-1}$. Such a generalization can
be easily be added to the method of Bray and Moore \cite{bm}.}.
This theorem immediately leads to the identity (compare \cite{II})
\begin{equation}\label{82}
\chi_l(\omega)= N^{-1} \sum _i\big \{[\beta(1-m_i^2)]^{-1} +
\chi_l-i \omega (\beta L_{ii})^{-1}-\chi_l(\omega)\big\}^{-1}
\end{equation}
and $\chi_l(\omega)$ is obtained as a solution of equation
(\ref{82}) satisfying $\omega \,\chi_l''(\omega)\geq 0$. In the
paramagnetic regime and in absence of  external magnetic fields
this equation leads to the analytic result
\begin{equation}\label{83}\fl
\chi_l(\omega)=\frac{\beta+\beta^{-1}- i \omega \beta^{-1} L^{-1}
}{2}+ \sqrt{\frac{(\beta+\beta^{-1}- i \omega \beta^{-1} L^{-1}
)^2}{4 }-1} \quad T\geq 1 \;,\; H_i=0
\end{equation}
where the Onsager coefficient $ L_{ii}=L(T)$ is temperature
dependent and given by
\begin{equation}\label{84}
L(T)= \exp(-\beta^2/2) \frac{1}{\sqrt{2 \pi}}\int\frac{
h\,\exp(-h^2/2) }{\sinh \beta h }\; \textrm{d}h\quad.
\end{equation}
using the equations (\ref{25}) and (\ref{50}). Note that for $T\gg
1$ equation (\ref{83}) reduces to $\chi_l= \beta(
1-i\omega\beta)^{-1}$ which implies a relaxation time proportional
to $ \beta $, as  is the case for the Korringa relaxation of
single magnetic moments dissolved in a metal. Note further that
the result (\ref{83}) partially agrees with the work of Kinzel and
Fischer \cite{kf,fh} who, however, have not obtained the
temperature dependence of $L(T)$. Near the spin glass temperature
both approaches agree and yield $\chi_l(\omega)=1 +i
\sqrt{|\omega|}\,$  at $ T=1$.

With reference to the definition of $ \Gamma$ \cite{II,III}
\begin{equation}\label{88}
\chi_l(\omega)\mid_{\omega\rightarrow\pm0}= \chi_l\pm i\Gamma
\end{equation}
results from equation (\ref{82}). Thus the quantities $\chi_l$ and
$\Gamma$, which enter formally in derivation the modified TAP
equations, are related to the real and the imaginary part of the
isolated susceptibility $\chi_l(\omega \rightarrow 0)$. Recalling
that there are frequently situations in physics where the isolated
and the isothermal susceptibility differ (for some general
discussion see \cite{fs}) this result gives some additional
insight into approach leading to the modified TAP equations
\cite{II,III}.
\begin{figure}
\begin{center}
\includegraphics[height=7.5cm]{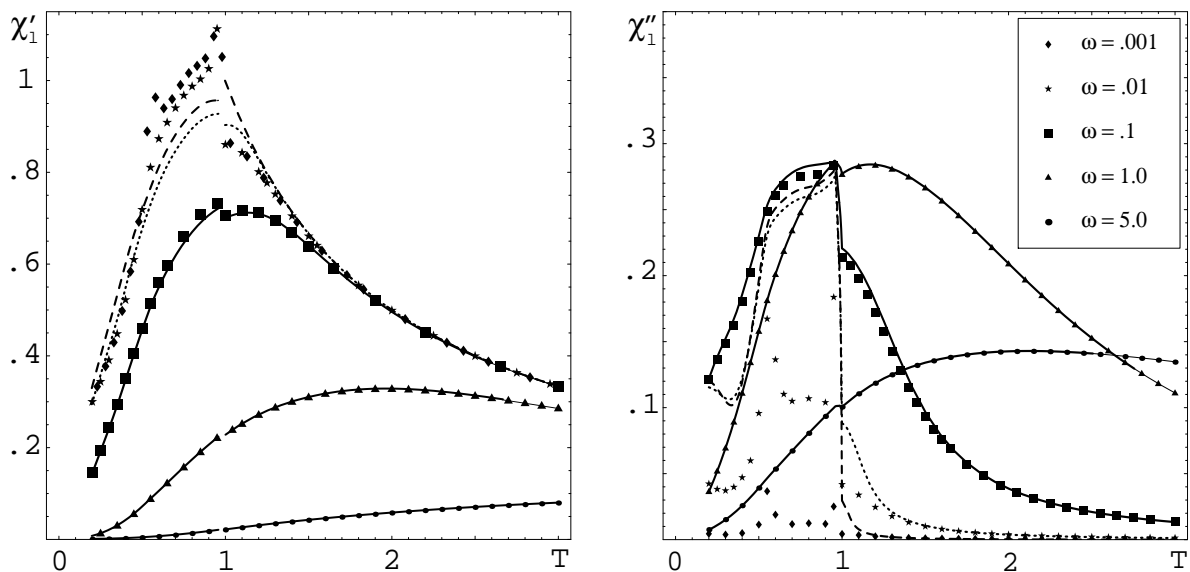}
\end{center}
\caption{\label{f34} Temperature dependence of the real part,
$\chi_l'(\omega) $, and the imaginary part, $\chi_l''(\omega) $,
of the local dynamic susceptibility $\chi_l(\omega) $ in zero
external field for different frequencies. The data points are
calculated by numerical inversion of equation (\ref{81}) for a
system of $ N=225 $ spins (compare text). The lines represent the
results calculated  from equation (\ref{82}) via the theorem of
Pastur. For $ T<1$ they are again based on the numerical data of
the $N=225$ spin system. For $ T>1$ the lines give the analytic
result of equation (\ref{83}) in the thermodynamic limit. The
dashed line, the dotted line and the full lines  correspond to
$\omega = 0.001$, to $\omega = 0.01$ and to $\omega= 0.1, 1.0 ,
5.0$ respectively. For $\omega = 0.001$ some data points exist
which are outside the plot-range of $\chi_l'(\omega) $. }
\end{figure}

In figure (\ref{f34}) both the analytical and the numerical
results for $\chi_l$  are presented for the zero external field
case. For the numerical parts again the lowest free energy state
of sample I from reference \cite{III} with $N=225$ is used.  The
two different methods lead to similar results but deviate
increasingly for small frequencies $\omega$. It is remarkable that
the results of figure (\ref{f34}) show some overall  similarity to
real, experimental data \cite{fh,young}. One of these features is
the frequency dependence of the cusp temperature of
$\chi_l'(T,\omega=const)$.  In the present approach the shift
results simply from the temperature dependence of the Onsager
coefficients.

The deviations appearing in figure (\ref{f34}) for small
frequencies are caused  by the finite-size of the system.
According to equation (\ref{231}) $\bi{L}$ is positive definite.
Thus the matrix $ \bi{A}= ( \beta \bi{L}) ^{1\over 2} $ is well
defined and $ \bchi^{-1} (\omega)$ can be rewritten as $
\bchi^{-1}(\omega)=\bi{A}^{-1}(\bi{A} \bchi ^{-1}\bi{A} - i\omega)
\bi{A}^{-1}$. Use of the diagonal representation of $\bi{A} \bchi
^{-1}\bi{A}$ leads to
\begin{equation}\label{85}\fl
\chi_l(\omega)=N^{-1} \sum_\alpha \langle
u_\alpha|\bi{A}^2|u_\alpha \rangle\;( \delta_\alpha -i
\omega)^{-1} \quad \textrm{where}\quad \bi{A} \bchi
^{-1}\bi{A}|u_\alpha \rangle=\delta_\alpha|u_\alpha \rangle\:
\end{equation}
which shows  that $\chi_l(\omega)$ can be written as a
superposition of Lorenzian functions with relaxation rates $
\delta_\alpha $ given by the eigenvalues of $\bi{A} \bchi
^{-1}\bi{A} $ . In the spin glass regime the minimum value
$\delta_{min}$ of all rates $ \delta_\alpha $ is small (the
numerical value for the data used in figure (\ref{f34}) is of the
order 0.03) for finite $N$ and tends to zero in the thermodynamic
limit \cite{II,III,bm}. For frequencies $\omega<\delta_{min} $ the
susceptibility $\chi_l(\omega)$ sensitively depends on the small
eigenvalues according to equation (\ref{85}) and thus finite-size
effects mainly show up for small $\omega$  for the data obtained
by direct matrix inversion. According to figure (\ref{f34}) the
deviations  are moderate for those results which use the theorem
of Pastur. This indicates that the use of this theorem for finite
$N$ smoothes partly out the finite-size effects.
\subsection{Relaxation function}
The local relaxation function  can be written as
\begin{equation}\label{86}
\Phi_l(t)= N^{-1} \Tr \bPhi(t) = N^{-1} \sum_\alpha \langle
u_\alpha|\bi{A}^2|u_\alpha
\rangle\;{\delta_\alpha}^{-1}\:\exp(-\delta_\alpha t)
\end{equation}
where  equation (\ref{29}) and the eigenvalue equation (\ref{85})
and were used. Again the superposition of the contributions
resulting from the different rates $ \delta_\alpha $ (or
relaxation times  $ \delta_\alpha^{-1} $ ) can be identified. The
local relaxation function satisfies $ \Phi_l(t)\leq
\Phi_l(0)=\chi_l.$ and thus is well behaved in the thermodynamic
limit even for the case when $ \delta_{min} $ tends to zero. For
vanishing external fields  $\Phi_l(t)$ can be given analytically
\begin{equation}\label{87}\fl
\Phi_l(t)= \frac{2}{\pi} \int_{-1}^1\,\textrm{d}x
\frac{\sqrt{1-x^2}}{\beta+\beta^{-1}-2x}
\exp\Big[-(1+\beta^2-2\beta x ) L(T) \Big]
 \quad T\geq 1 \;,\; H_i=0
\end{equation}
in the paramagnetic regime \cite{sk}. As already noted, the
temperature dependence of  $L(T)$ differs from \cite{sk,szamel}
and is  given by equation (\ref{84}).
\begin{figure}
\begin{center}
\includegraphics[height=8cm]{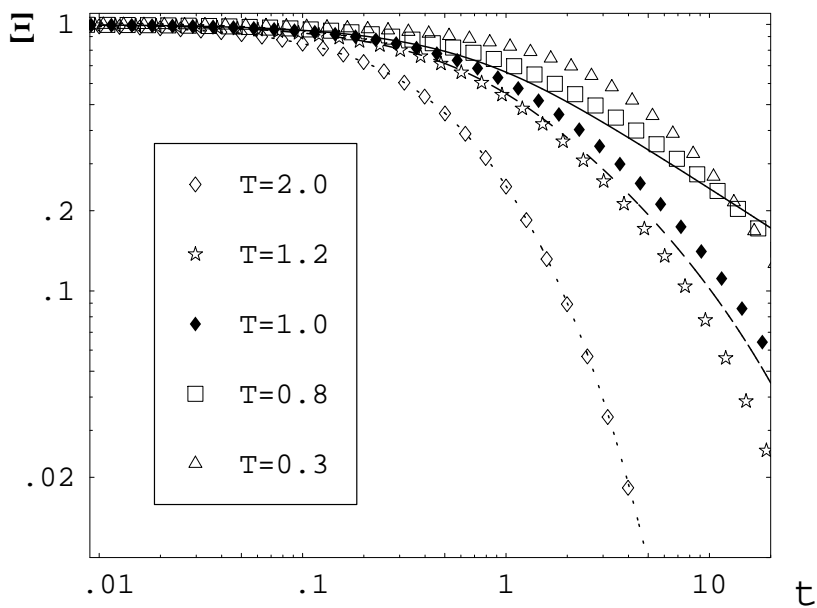}
\end{center}
\caption{\label{f5} Time $t$ dependence of the reduced local
relaxation function $ \Xi(t)= \Phi_l(t)/\Phi_l(0)$ in a double log
scale for different temperatures $T$ in zero external field. The
data points correspond to the numerical results of a sample with
$N=225$ spins. The  lines represent the  results in the
thermodynamic limit given by equation (\ref{87}). The full, the
dashed and the dotted line correspond to $ T=1.0 ,1.2$ , and $2.0$
respectively.}
\end{figure}

In figure (\ref{f5}) the numerical results for the local
relaxation function $\Phi_l(t)$ are plotted (again for the state
with lowest free energy of sample I of reference \cite{III})
together the analytic results (\ref{87}). The plot clearly
exhibits the slowing down at the spin glass temperature and the
presence of the slow dynamics in the spin glass regime. Comparing
the numerical and the analytical results above the spin glass
temperature the deviations increase for the long time behavior
when approaching the spin glass regime. As discussed already
above, these finite-size effects result from the finite value of $
\delta_{min }$ and limits the numerical results to the region $
t\ll \delta_{min} \approx 30 $ near and below the spin glass
temperature.
\section{Conclusions}
Two questions are studied  from a more general point of view  in
the present work. First of all the linear response theory for an
arbitrary Ising model which is microscopically and weakly coupled
to a bath is investigated using the theory of Mori and applying
the standard approximations. This approach seems to be natural and
conservative and could have been carried out some decades ago.
Nevertheless the results are remarkable, as the entire dynamical
response of any Ising model is completely determined by the
internal field distribution function, by the static isothermal
susceptibility and - as the only characteristic feature of bath -
by a bath dynamic susceptibility. Due to the simple structure this
part of the present work may be of interest for other models than
the SK model.

The relation to former approaches has been worked out in some
detail at the end of section 2. It is the use of microscopically
unjustified transition rates in these master-equations which
causes the differences to the present investigation.

The second question is exclusively related to the SK model and
represents  the study of the internal field distribution function
showing how this  function is related to the solutions of the TAP
equations. The relationship obtained  opens the possibility of
explicitly calculating further quantities of interest for the SK
model, such as the inelastic-neutron-scattering cross section
\cite{ttcs}.

A further tool to obtain the final results for the dynamic linear
response functions of the SK model is the explicit knowledge of
the solutions of the modified TAP equations. This again
demonstrates the importance of the modified TAP approach.

The present work is limited to the dynamics in linear
approximation near the thermodynamic equilibrium. It is, however,
well known that in the physics of spin glasses nonlinear dynamical
effects and the linear response  in out of equilibrium situations
are of great importance \cite{young}. Thus an extension of the
present microscopic approach to nonlinear dynamics  would be of
great use to treat these effects on a well-founded bases. Such an
approach based on the theory of Robertson  or on the theory of
Nakajima-Zwanzig (see e.g \cite{fs}) is in progress and will be
published separately \cite{V}.

\ack Interesting discussions with G. Sauermann are acknowledged.
%\appendix
%\setcounter{section}{1}

\end{document}